\documentclass[twocolumn,prl,showpacs,preprintnumbers,amsmath,amssymb,superscriptaddress]{revtex4}

\usepackage{graphicx}
\begin{document}

\title{Evaluation of conduction eigenchannels of an adatom probed by an STM tip}

\author{Martyna Polok}
\email{martyna.polok@physik.uni-halle.de}
\affiliation{Institut f\"ur Physik, Martin-Luther-Universit\"at Halle-Wittenberg, D-06099 Halle, Germany}
\author{Dmitry V.~Fedorov}
\affiliation{Institut f\"ur Physik, Martin-Luther-Universit\"at Halle-Wittenberg, D-06099 Halle, Germany}
\author{Alexei Bagrets}
\affiliation{Steinbuch Centre for Computing and
 Institute of Nanotechnology, Karlsruhe Institute of Technology,
Hermann-von-Helmholtz-Platz 1, 76344 Eggenstein-Leopoldshafen}
\author{Peter Zahn}
\affiliation{Institut f\"ur Physik, Martin-Luther-Universit\"at Halle-Wittenberg, D-06099 Halle, Germany}
\author{Ingrid Mertig}
\affiliation{Institut f\"ur Physik, Martin-Luther-Universit\"at Halle-Wittenberg, D-06099 Halle, Germany}
\affiliation{Max-Planck-Institut f\"ur Mikrostrukturphysik, Weinberg 2, D-06120 Halle, Germany}

\date{\today}
\begin{abstract}
Ballistic conductance through a single atom adsorbed on a metallic surface and probed by a scanning tunneling microscope (STM) tip can be decomposed into eigenchannel contributions, which can be potentially obtained from shot noise measurements. Our density functional theory calculations provide evidence that transmission probabilities of these eigenchannels encode information on the modifications of the adatom's local density of states caused by its interaction with the STM tip. In the case of open shell atoms, this can be revealed in nonmonotonic behavior of the eigenchannel's transmissions as a function of the tip-adatom separation.
\end{abstract}
\maketitle
%
A deep understanding of the transport properties of atomic size systems is of substantial importance and has received considerable interest, spurred, in particular, by the possible applications of nanoscale conductors in future electronic device technologies. Thanks to advances in scanning tunneling microscopy, it is already possible not only to probe the electronic and magnetic structure of surfaces, but also to explore conduction properties of atomic size systems, such as one-dimensional wires~\cite{Ohnis98},
individual organic molecules~\cite{Repp2010,Venkataraman2009,Schull2010,Tautz2010,Wulfhekel2011},
atomic sized contacts~\cite{Scheer1997},
or even single atoms~\cite{Wiesendanger2010,neel07,neel09,Brune09,tao10,tao09}.
With regard to the latest case, recent experimental observations by Berndt and coworkers~\cite{limot05,neel07} revealed
a qualitative difference in the conductance behavior depending on whether the STM tip was approaching a clean surface or a single adsorbed atom. Contrary to the sudden and unpredictable jump in the conductance for the clean surface,
a smooth and fully reversible transition from the tunneling to the contact regime was observed for Cu(111) and Ag(111) surfaces decorated with Cu, Ag, or Co adatoms. Here an increased bonding of the adatom to the surface, due to a dipolar contribution caused by the redistribution of the surface charge, was believed to play a crucial role \cite{limot05}.

%
%
The case of magnetic adatoms deposited on nonmagnetic substrates is especially attractive \cite{Kern2002} since the characteristic Kondo temperature ($T_K$)
has been observed to be sensitive to modifications of the local density of states (LDOS) at the adatom owing to hybridization of its atomic wave functions with the STM tip \cite{neel07}.
Here we offer yet another LDOS probe. The conductance of the atom-sized contact can be decomposed into contributions
from the individual transport eigenchannels.
In this communication we argue that the transmission probabilities
of these eigenchannels, which in principle can be obtained from the shot noise measurements \cite{Tal2008}, could serve, in addition to $T_K$, as a sensitive tool to probe the modifications in the local electronic structure of the (magnetic) adatom interacting with the (spin-polarized) STM tip.

\begin{figure}
\includegraphics[width=0.25\textwidth]{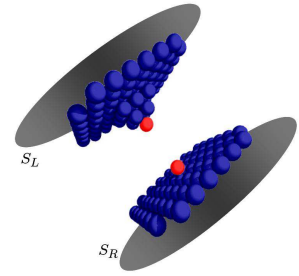}
\caption{%
A schematic STM model.
The red atoms indicate positions where Cu/Co atoms were introduced to the system.
The gray areas correspond to atomic planes used for the evaluation of the conductance, situated four monolayers above the tip ($S_L$) and below the adsorbed atom ($S_R$).
}
\label{stm}
\end{figure}
%
We present an {\it ab initio} study of the tunneling conductance for an STM tip approaching a Cu$(001)$ surface decorated with a single Cu or Co adatom (Fig.~1).
%
Our calculations are based on density functional theory \cite{mp}
within the screened Korringa-Kohn-Rostoker (KKR) Green's function formalism \cite{papan02a}.
The potentials were assumed to be spherically symmetric around each atom.
Nonetheless, the full charge density, rather than its spherically symmetric part, was taken into account.
The angular momentum cut-off of
$l_{\rm{max}}=3$
was used.
Employing an embedding technique for the Green's functions, we treat the STM tip and the adsorbed atom as an impurity-like cluster,
which perturbs the otherwise perfect structure of two semi-infinite crystalline leads separated by the vacuum region \cite{bagrets06}.
%
The zero bias conductance was calculated using the Kubo linear response theory in the formulation of Baranger and Stone \cite{baranger89}. Within the mixed site and orbital-momentum representation of the KKR method, pairwise contributions to the conductance have been summed up in real space between two atomic planes, $S_L$ and $S_R$ in Fig.~1, located in the leads perpendicular to the $z$ direction in which the electrical current flows \cite{KKR-conductance}.

\begin{figure}
\includegraphics[width=.4\textwidth]{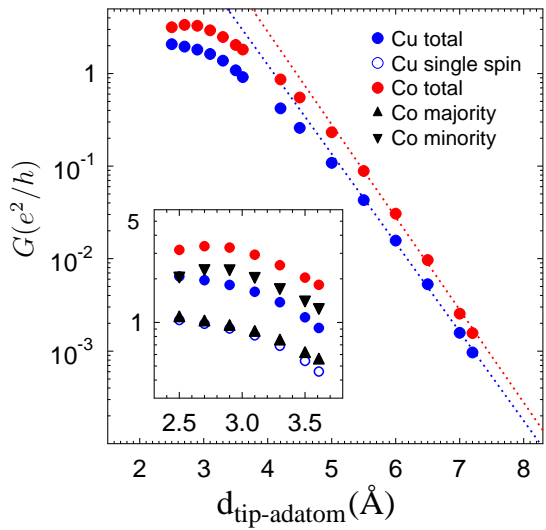}
\caption{Conductance as a function of the tip-adatom
separation. The inset shows a close up of the conductance for small tip-adatom separation including the spin-polarized contributions.}
\label{cond}
\end{figure}

%
In Fig.~2 we present results for conductance $G(d)$ versus tip-adatom separation $d_{tip-adatom}$ computed for two setups: (i) a nonmagnetic case with a Cu adatom on Cu(001) surface approached by a Cu tip (blue points); and (ii) a magnetic Co atom deposited on the Cu(001) surface probed by a modified Cu tip with a Co apex atom (red points). For the latter we assumed a parallel alignment of the Co magnetic moments.
As the STM tip is approaching the surface the tunneling conductance increases exponentially and thereafter smoothly evolves into the contact regime, similarly to the experimental observation~\cite{limot05,neel07}.
However, depending on the chemical nature of the system, with decreasing $d_{tip-adatom}$ the transition from the tunneling to contact regime emerges in a different way.
In the case of the open shell magnetic Co adatom (assuming we are above $T_K \sim 100$~K), the conductance experiences a strong increase in the minority spin channel (inset of Fig.~2).
%
From the exponential behavior at large distances ($\ge 5.5$~\AA) we have extracted
the work functions for the systems: 5.0~eV for the Cu adatom, and 5.8~eV for the Co adatom, which are typical for metals. The first value agrees well with the tunneling barrier height of 5.3~eV found in STM experiments by Berndt {\it et al.}~\cite{limot05}.
For small tip-adatom separation the conductance saturates to values $\simeq 2.0~e^2/h$ for the Cu system, and $\simeq 3.0~e^2/h$ for the Co system, which are typical for metallic atomic point-contacts.

\begin{figure}
\includegraphics[width=.4\textwidth]{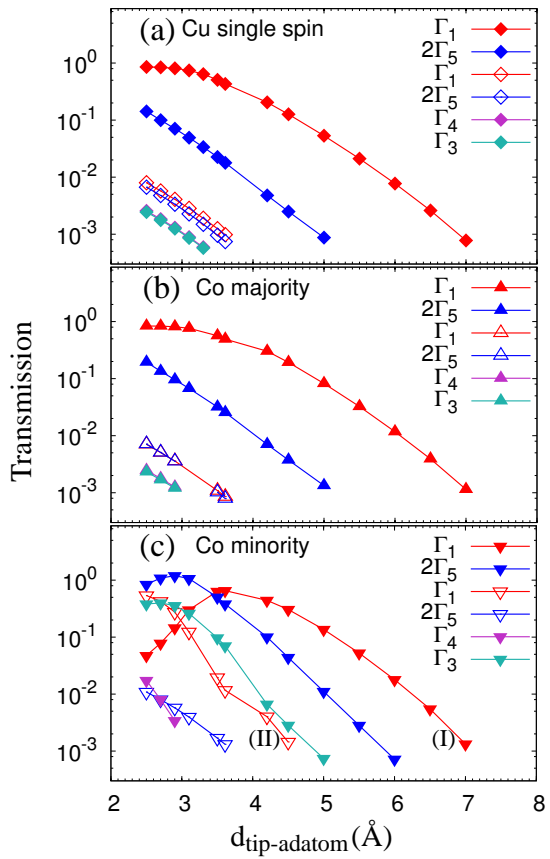}
\caption{Transmissions of the conduction eigenchannels
as a function of separation between the STM tip and the adatom:
(a) a nonmagnetic Cu system; (b) majority spin channel of the magnetic system
composed of Cu surface with a Co adatom probed by the Cu tip with a
Co apex atom, and (c) the corresponding minority spin channel.}
\label{tr}
\end{figure}

\begin{figure*}
\includegraphics[width=1.0\textwidth]{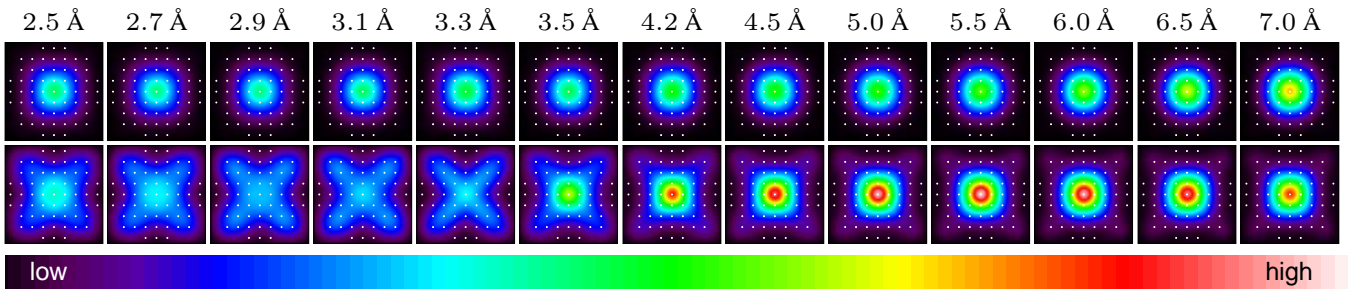}
\caption{Spin decomposed current density in the $S_R$ plane (see Fig.~1) for majority (top row) and minority (bottom row) spin channels (tip-adatom separation ranging from  2.5 \r{A} to 7.0 \r{A}). White dots represent the atomic sites.}
\label{curr}
\end{figure*}

%
According to the Landauer approach a microscopic insight into
the conductance is possible by analyzing its decomposition into independent conduction modes (eigenchannels). Namely,
$G =  G_0\mathrm{Tr} [\tau\tau^{\dagger}] = G_0\sum_n {T_n}$,
where $G_0=2e^2/h$ is the conductance quantum and the eigenchannels' transmissions $T_n$  are defined as eigenvalues of the matrix $\tau\tau^{\dagger}$.
The matrix elements $\tau_{mn}$ are transmission amplitudes for scattering of incoming waves $m$ into outgoing waves $n$ in the opposite lead. In the case of a magnetic system, the above formulation is applied to each spin channel separately, neglecting spin-orbit coupling.
Experimental and theoretical studies \cite{Scheer1997, Scheer1998, Cuevas1997} of atomic wires and atomic size contacts revealed a close relation between the conduction channels and the chemical nature of atoms, in particular, the occupation of their valence atomic orbitals.
Results of our calculations for the contact regime are in agreement with common expectations.
Namely, due to the open shell structure of transition metal atoms (Co in our case),
one expects several conduction channels being partially open and conductance
values above $G_0$ --- a situation different from noble metal
single atomic wires where the conductance is close to $G_0$,
which is usually dominated by one channel per spin with almost perfect transmission.

%
We have performed a quantitative analysis of the eigenchannel contributions to conductance
using a recently proposed approach \cite{bagrets06,bagrets07}, which employs a mapping of the
transmission matrix $\tau\tau^{\dagger}$ onto a site and orbital-momentum KKR basis set.
Within this approach we analyze the symmetry of the solutions of the eigenvalue problem for $\tau\tau^{\dagger}$.
We adopt a classification of eigenchannels in terms of the internal degrees of freedom of the selected atom placed at the bottleneck of the STM tunnel junction, i.e.\ the adatom.
Such a classification emerges naturally if one considers the wave functions of eigenchannels to be projected onto the adatom's valence orbitals, which are expected to play a decisive role in the conduction process \cite{Scheer1998}.
With regard to the $C_{4v}$ symmetry of the  fcc (001) surface, one is thus
able to distinguish conduction modes corresponding to the one-dimensional
$\Gamma_1$ (including $s$, $p_z$, $d_{z^2}$ contributions), $\Gamma_3$ ($d_{x^2-y^2}$) and
$\Gamma_{4}$ ($d_{xy}$) irreducible representations, and to the two-dimensional
$\Gamma_{5}$~($\{p_x$, $d_{xz}\}, \{p_y$, $d_{yz}\}$) representation.
%
%

Employing a symmetry analysis, we have found out that
in the case of the Cu adatom probed by the Cu tip, Fig.~3a, only one $\Gamma_1$ eigenchannel contributes considerably to the conductance and dominates the transmission for any $d_{tip-adatom}$.
 Other contributions to the total transmission are orders of magnitude smaller.
  Similarly, there exists only one dominant $\Gamma_1$ eigenchannel for the majority spin states of a magnetic Co adatom probed by the modified spin-polarized STM tip, Fig.~3b.
    In contrast, for the Co minority spin states at smaller $d_{tip-adatom}$ almost all symmetry contributions play a significant role, since the partially occupied $d$ orbitals start to overlap, Fig.~3c. Around $d_{tip-adatom}\approx$ 3\AA\ there exist five eigenchannels: two $\Gamma_1$ channels, the double degenerate $\Gamma_5$ channel, and the $\Gamma_3$ channel.
  Moreover, and most interestingly, the transmission of these eigenchannels does not always change monotonically with $d_{tip-adatom}$.
In the following we show that such a behavior is ultimately related to the variations of the LDOS at the Co adatom induced by the interaction with the STM tip.

%
Simultaneously to the change of the eigenchannels' contributions caused by the approaching tip, we also analyzed the behavior of the complementary information, namely a distribution of the current density in the $S_R$ plane, Fig.~\ref{curr}.
With a smaller $d_{tip-adatom}$ the current density for the majority spin channel defocuses but otherwise, as expected, remains circularly symmetric about the center of $S_R$. This symmetry is compatible with the $\Gamma_1$ representation.
In contrast, the current density for the minority spin states changes its nature completely and undergoes a strong modulation reflecting the role of the $d$ orbitals. The distance range for which the current density changes its character corresponds to $d_{tip-adatom} \lesssim 3.5$~\AA, where the transmission of the minority spin electrons of Co is dominated by the $\Gamma_5$ channel (Fig.~\ref{tr}c).

%
Understanding of the above results is possible if we recall the projection of the eigenchannels' wave functions onto the adatom's orbitals.
The $\Gamma_1$ channel of Cu and of the Co majority states is naturally attributed to valence 4s orbitals, while
the double degenerate $\Gamma_5$ channel can be related to the valence $p_x$ and $p_y$ orbitals.
Since for Cu and Co majority states the weight of $p$ states at $E_F$ is very low, their contribution to
conductance in the contact regime is factor 10 smaller than $\Gamma_1$.

The behavior of the Co minority spin eigenchannels as a function of $d_{tip-adatom}$ (Fig.~3c) can be explained by the LDOS at the adatom (Fig.~5).
In contrast to the majority spin d-states of Co, which
are filled, the minority spin states are partially occupied (Fig.5) which
gives rise to a magnetic moment of 1.8 $\mu_B$ at the Co adatom. Moreover,
we notice that at the contact point  between the adatom and the STM tip, $d_{tip-adatom} \simeq 2.5$~\AA,
each minority $d$ orbital contribution is characterized by a double-peak structure (Fig.~5a). Because of a
mutual interaction between the Co tip and Co adatom a dimer with bonding and antibonding states is formed.
As the STM tip is being retracted from the surface, the overlap between atomic orbitals is gradually reduced (Fig.~5b), thus the splitting between bonding and antibonding states diminishes leading to the shift of minority  $d$ states towards $E_F$ (Fig.~5c) and an increase in the corresponding LDOS (Fig.~5d-f).

As one can see in Fig.~5a-c, the $d_{xy}$ states are almost fully occupied, and therefore their contribution to conductance ($\Gamma_4$ channel) is negligible. The other four minority $d$ orbitals are partially occupied and contribute to the conductance in the contact regime.
The $d_{x^2-y^2}$ states are located in the close vicinity to $E_F$, giving rise to the $\Gamma_3$ channel.
With increasing $d_{tip-adatom}$ the $d_{x^2-y^2}$ states accumulate at $E_F$ (Fig.~5f) and support transmission of the  $\Gamma_3$ channel until it finally starts to decay exponentially at larger distances.

A similar explanation can be provided for the minority $\Gamma_5$ channel of Co.
In the contact regime, this channel is strongly influenced by the minority $d_{xz}$ and $d_{yz}$ states of the Co adatom.
Due to the weakening of the Co-Co bond with increasing of the tip-adatom separation, the $d_{xz}$, $d_{yz}$ LDOS peak located at $0.3$~eV below $E_F$ for small $d_{tip-adatom}$ (Fig.~5a), approaches the Fermi level leading to an increase of the LDOS($E_F$) (Fig.~5e), and therefore to the increase in the conductance.
The exponential decay later for larger $d_{tip-adatom}$ is driven by $p_x$, $p_y$ contributions to this channel, in analogy to the case of the majority states.

Finally, two minority $\Gamma_1$ channels, $\Gamma_1$(I) and $\Gamma_1$(II) in Fig.~3c, can be interpreted as superpositions of the $4s$ and $3d_{z^2}$ states, which share the same in-plane symmetry.
At the contact point the $\Gamma_1$(I) channel (filled red triangles) is nearly closed and thus it is projected mainly onto the states with $d_{z^2}$ symmetry, whose weight at $E_F$ is initially low (Fig.~5a).
The $\Gamma_1$(II) channel (open red triangles) has a mainly $s$ like character and is open at the contact point.
When the tip is retracted from the adatom, the antibonding $d_{z^2}$ resonance approaches $E_F$ (Fig.~5a-c,d), which leads to the opening of the previously closed $\Gamma_1$(I) channel. Above the crossing point, $d \simeq 3.0$\AA, the two channels are hybrids of both orbitals, $s$ and $d_{z^2}$.
The $\Gamma_1$(I) channel saturates to a transmission value of almost 1.0 due to the $d_{z^2}$ resonance pinned to the Fermi level. For larger tip-adatom separations this channel acquires more $s$ weight, and decays exponentially similarly to the majority spin $\Gamma_1$ channel (Fig.~5b). Simultaneously, the $\Gamma_1$(II) channel acquires more $d_{z^2}$ character and decays much faster due to the reduced overlap between the $d$ orbitals localized at the adatom and the tip.

%

Summarizing, we have demonstrated that modifications of the adatom's electronic structure
due to interactions with an STM tip can result in nonmonotonic
behavior of
eigenchannel's transmission probabilities as a function of the tip-adatom separation.
An experimental confirmation of this result is expected from the analysis of
shot noise level measurements. Moreover, the eigenchannel decomposition
procedure presented here may provide a deeper microscopic and material-specific insight into the transport
phenomena when applied to various STM related studies.

\begin{figure}
\includegraphics[width=.49\textwidth]{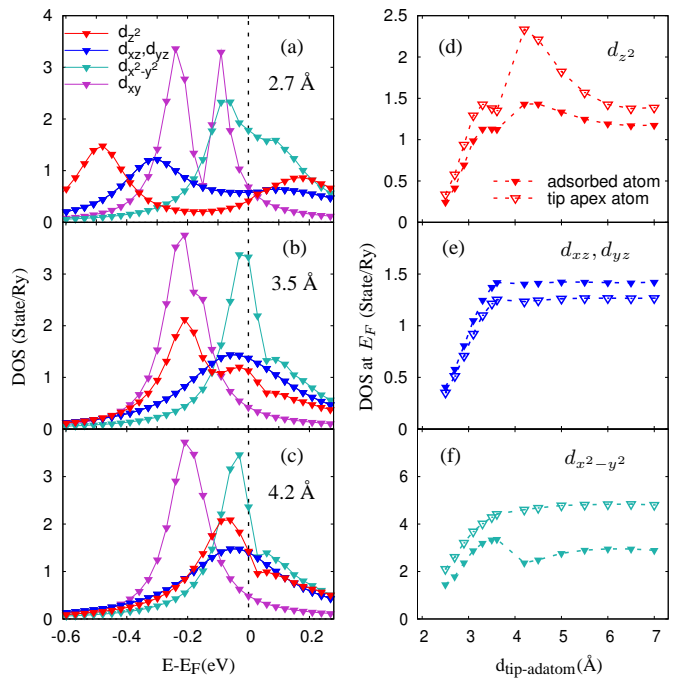}
\caption{
Left: evolution of the minority spin local
density of states (LDOS) at the Co adatom
as a function of its separation $d$
to the STM tip, (a) $d = 2.7$~\AA, (b) $d = 3.5$~\AA, (c) $d = 4.2$~\AA.
The contributions of the five $d$ orbitals are shown.
Right: symmetry decomposed
LDOS computed at the Fermi level (d-f), the contributions at the adatom (closed symbols) are
compared with those at the tip apex atom (open symbols).
}
\end{figure}

This work was supported by the DFG (Deutsche Forschungsgemeinschaft),
under projects SFB762 and SPP1243, and the
"Forschungsnetzwerk Nanostrukturierte Materialien" of the state Sachsen-Anhalt.

\end{document}